\documentclass[11pt,a4paper]{article}

\usepackage[left=2.2cm,right=2.0cm,top=2.1cm,bottom=2.05cm,footskip=0.75cm]{geometry}
\usepackage{iftex}
\ifPDFTeX
  \usepackage[T1]{fontenc}
  \usepackage{newtxtext}
\else
  \usepackage{fontspec}
\fi
\usepackage[colorlinks=true,citecolor=blue,urlcolor=blue,linkcolor=black]{hyperref}
\usepackage{amsfonts,amsmath,amssymb,amsthm}
\ifPDFTeX
  \usepackage{newtxmath}
\fi
\usepackage{derivative}
\usepackage{lineno}
\usepackage{afterpage}

\usepackage{bm}
\usepackage{xurl}
\usepackage{srcltx}
\usepackage{graphicx} 
\usepackage{caption}
\usepackage{subcaption}
\usepackage{titlesec}
\usepackage{microtype}
\usepackage{enumerate}
\usepackage{xcolor}
\usepackage{here}
\usepackage{caption}
\usepackage[ruled,linesnumbered]{algorithm2e}

\usepackage[affil-it]{authblk}
\usepackage[english]{babel}
\usepackage{lineno}
\usepackage{booktabs}

\numberwithin{equation}{section}

\titleformat{\section}{\bfseries\fontsize{13pt}{15pt}\selectfont}{}{0pt}{}
\titlespacing*{\section}{0pt}{18pt}{4pt}
\titleformat{\subsection}{\bfseries\fontsize{11pt}{13pt}\selectfont}{}{0pt}{}
\titlespacing*{\subsection}{0pt}{14pt}{3pt}

\DeclareCaptionFont{referencecaption}{\fontsize{9.5pt}{12.9pt}\selectfont}
\makeatletter
\renewcommand{\@maketitle}{%
  \newpage
  \null
  \vspace*{-1.25em}%
  \begin{center}%
    {\bfseries\fontsize{18pt}{21pt}\selectfont \@title\par}%
    \vspace{13pt}%
    {\lineskip .5em\begin{tabular}[t]{c}\@author\end{tabular}\par}%
  \end{center}%
  \vspace{10pt}%
}
\makeatother


\theoremstyle{definition}

\theoremstyle{remark}

\usepackage{rotating}
\usepackage{tablefootnote}
\usepackage{url}
\usepackage{pifont}

\usepackage{chngcntr} 
\counterwithout{equation}{section} 

\usepackage{zref-xr,zref-user}

\begin{document}

\title{OceanMind: A multi-agent AI system for ocean diagnosis
}

\author[1,2,+]{Fan Zhang}
\author[2,3,+]{Weicong Cheng}
\author[1]{Yuheng Chen}
\author[2,3]{Hiuseut Kung}
\author[2,3]{Ying Zhang}
\author[2,3]{Aixi Han}
\author[2,3]{\\Quanjia Zhong}
\author[1,2,*]{Can Yang}
\author[1,2,3,*]{Jianping Gan}

\affil[1]{Department of Mathematics, The Hong Kong University of Science and Technology, Hong Kong SAR, China.}
\affil[2]{Center for Ocean Research in Hong Kong and Macau,  The Hong Kong University of Science and Technology, Hong Kong SAR, China.}
\affil[3]{Department of Ocean Science, The Hong Kong University of Science and Technology, Hong Kong SAR, China.}
\affil[+]{These authors contributed equally to this work.}
\affil[*]{Correspondence: magan@ust.hk, macyang@ust.hk}

\date{}

\maketitle

\section*{Abstract} \label{abstract}

Time-dependent, three-dimensional (3D) oceanic multi-variables define coherent states of the evolving ocean to facilitate ocean diagnosis and advance ocean science to better inform environmental and hazard management. However, extracting quantitative evidence from these variables requires substantial and complex analytical effort and time. Here, we introduce OceanMind, a multi-agent AI system that directly couples large language models (LLMs) with comprehensive time-dependent 3D ocean states for swift and effective diagnosis. OceanMind organizes the analytical process into four coordinated complexity stages: Query Routing, Skill-based Planning, Tool Execution, and Evidence-based Summary Generation. Across these stages, specialized agents collaborate to interpret user requests, construct and execute multi-step computational workflows, and synthesize quantitative evidence. To ensure reliable computational workflow construction, 63 reusable ocean-specific analysis skills serve as procedural manuals that guide the LLM agent in selecting appropriate data, conducting diagnostics, and applying analytical tools for each objective request. With reflection and replanning mechanisms that use execution feedback to repair invalid plans, OceanMind ensures reliable analysis workflows across diverse needs. On a comprehensive benchmark of 240 computational-workflow queries spanning the four complexity stages, OceanMind outperformed general ReAct agents that had the same registered tool pool, achieving relative improvements of 41.2\% in effectiveness and 21.5\% in efficiency. Beyond the benchmark, OceanMind demonstrated its capabilities in reproducing published oceanic diagnostics, validating hypotheses, and supporting environmental decision-making over global oceans. Overall, OceanMind advances LLMs by integrating the LLMs with time-dependent 3D ocean analysis, enabling scientific interpretation and enhancing formulation of environmental policies based on quantitative ocean evidence.

\section*{Introduction} \label{intro}

Understanding how ocean conditions change is fundamental to advancing ocean science and informing environmental and hazard management~\cite{visbeck2018ocean,stammer2019ocean,kirchhoff2013actionable}. Diagnosing ocean-state changes from regional to global scales requires a coherent view of the evolving ocean across horizontal space, depth, and time~\cite{sun2023subsurfaceheatwaves}. Time-dependent, three-dimensional (3D) ocean state estimates provide the required view by resolving physical and biogeochemical conditions within a coherent spatiotemporal framework. However, comprehensive data do not automatically translate into scientific understanding or evidence. The effective use of diverse analytical tools demands computational skills and considerable effort to learn their requirements~\cite{m2024augmenting}. Experts must further integrate these tools into coherent workflows encompassing data processing, visualization, hypothesis testing, and interpretation~\cite{wang2023scientific,almansi2019oceanspy,stern2022pangeo}. These demands make these complex analyses even less accessible to non-specialists. Therefore, the scientifically and socially valuable information inherent within ocean states requires a much more convenient means of exploitation.

Large language models (LLMs) offer a promising route through this bottleneck~\cite{wang2023scientific,boiko2023autonomous,m2024augmenting}. However, using LLMs to automate the analysis of ocean states is challenging. Such analysis requires a system to interpret the user's intent, retrieve relevant context from existing information on the web, extract quantitative evidence from ocean states, and, finally, present the synthesized results transparently. General purpose~\cite{roberts2020much,achiam2023gpt,bai2023qwen} and domain-adapted LLMs~\cite{bi2024oceangpt,jin2025search} cannot directly analyze ocean states to meet user needs. LLMs rely on publicly available information and can produce quantitative and qualitative descriptions of ocean states that sound plausible but are actually inaccurate.~\cite{ji2023survey,achiam2023gpt,farquhar2024hallucinations}. OceanAI is an important early effort to use LLMs to extract quantitative evidence directly from selected ocean data products~\cite{chen2025oceanai}. However, the analytical demands of comprehensive time-dependent 3D ocean state estimates are substantially broader than OceanAI's current scope. Vertically, OceanAI's supported datasets are confined to limited variables, lack spatiotemporal coverage, and do not resolve the 3D physical-biogeochemical nature of the ocean~\cite{noaaOisst,noaaSeaLevelTrends,noaaCoopsWaterLevels}. In terms of disciplinary coverage, OceanAI's available data variables are limited to physical quantities and exclude biogeochemical fields. Within this limited scope, OceanAI uses a single coordinating agent to invoke predefined functions for data retrieval, basic statistics, and spatial or time-series visualization.  Being able to link multi-variables and physical-biogeochemical processes is critical to accurately diagnose and forecast the oceanic environment to provide coherent and holistic oceanic states, which is a requirement that OceanAI cannot fulfill.

\begin{figure}[htp]
  \centering
  \includegraphics[width=0.96\textwidth,keepaspectratio]{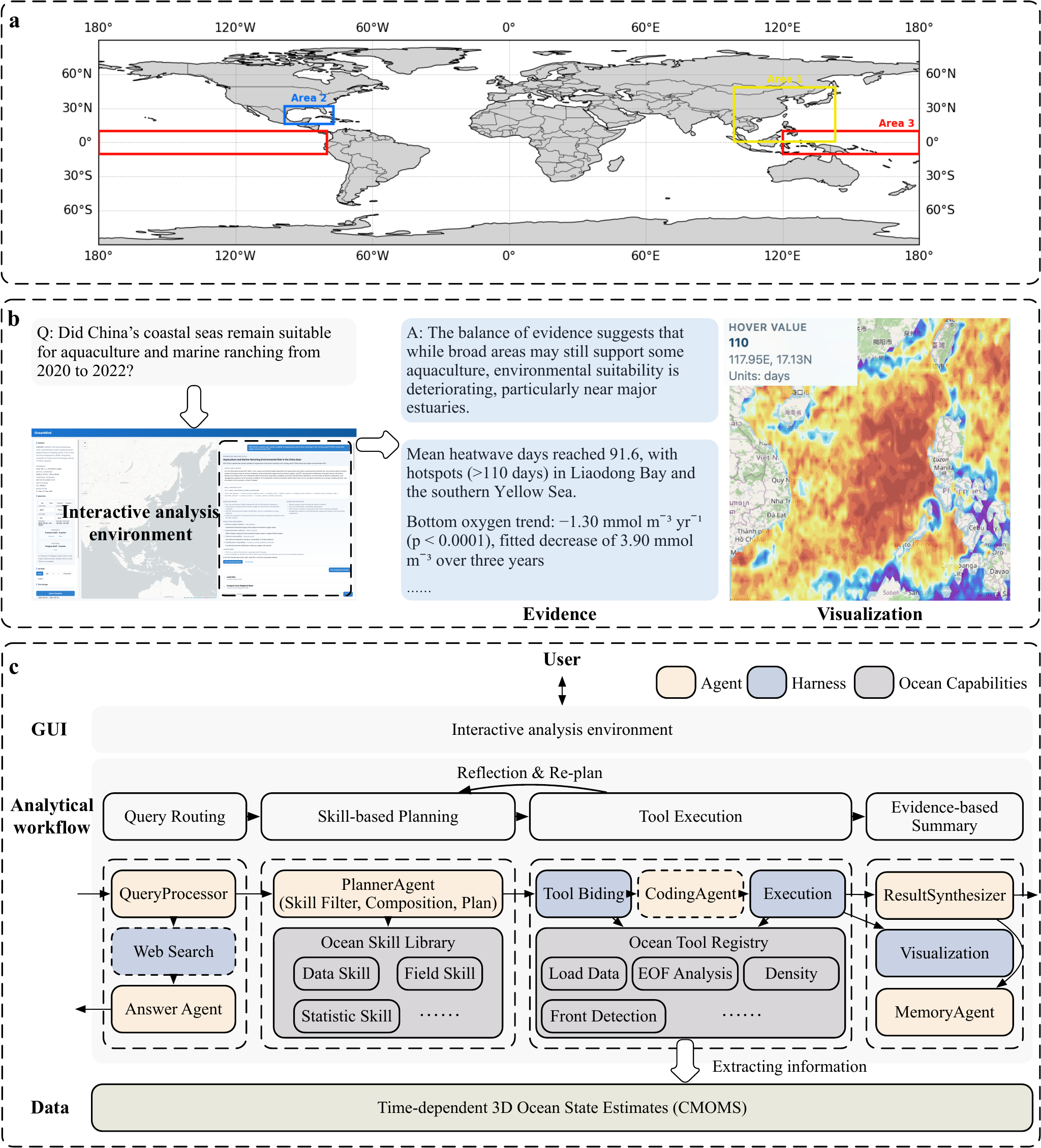}
  \caption{\textbf{Overview of OceanMind.}
\textbf{a.}  Geographic domains used to evaluate OceanMind. Area 1 covered the China Seas and adjacent western Pacific using CMOMS, whereas Areas 2 and 3 represented the Gulf of Mexico region and tropical Pacific using global CMEMS products, respectively.
\textbf{b.} Example of an interactive OceanMind analysis evaluating the suitability of China’s coastal seas for aquaculture and marine ranching. The graphical user interface (GUI) presented the analysis, quantitative evidence, spatial visualization, and synthesized response. 
\textbf{c.} Flow chart of the multi-agent architecture of OceanMind. User requests are processed through four coordinated stages: Query Routing, Skill-based Planning, Tool Execution, and Evidence-based Summary Generation. The QueryProcessor interprets each request and routes the request to an Internet-based response path or to computational analysis. For computational requests, the PlannerAgent selects and composes reusable skills from the Ocean Skill Library to construct a multi-step workflow. The execution harness binds appropriate tools, invokes the CodingAgent when needed, and performs computations from the Ocean Tool Registry on time-dependent 3D ocean state estimates. The ResultSynthesizer integrates the resulting evidence, and the Visualization module and MemoryAgent support result inspection and continued interaction. Reflection and replanning return execution feedback to the PlannerAgent when the workflow needs to be corrected.
  }
  \label{fig:overview}
\end{figure}

In this study, we introduce OceanMind, a multi-agent AI system that directly couples LLMs with comprehensive time-dependent 3D multi-variables and physical-biogeochemical processes to automate analyses for hindcasting and forecasting ocean states (Fig.~\ref{fig:overview}). To enable this end-to-end analytical process, OceanMind organizes these functions into four coordinated stages: Query Routing interprets user intent and identifies the required information sources; Skill-based Planning constructs a multi-step computational workflow that specifies how to extract the required information from the estimates; {Tool Execution} retrieves relevant Internet context when needed and extracts quantitative evidence from ocean state estimates; {Evidence-based Summary Generation} integrates these outputs and presents the synthesized results transparently. Specialized LLM agents perform distinct functions across these stages and together form OceanMind's collaborative multi-agent architecture. Among the four stages, {Skill-based Planning} is particularly challenging because diverse analytical requests require different computational workflows rather than a common sequence of operations. Complex requests might involve long chains of interdependent tool calls, with some mechanism-ranking and evidence-synthesis tasks requiring more than 20 calls. To address this challenge, OceanMind incorporates 63 reusable ocean-specific analytical skills that serve as procedural manuals for the PlannerAgent in Skill-based Planning. For each type of analytical request, a skill guides the PlannerAgent in resolving the required data selections, choosing appropriate diagnostics, and coordinating compatible tools to produce the expected evidence. To improve the reliability of these plans, OceanMind further incorporates reflection and replanning mechanisms. When execution reveals a failed operation, invalid intermediate result, or unresolved dependency, the feedback is returned to the PlannerAgent to repair the workflow. By doing so, we endow OceanMind with the ability to conduct a reliable end-to-end analytical process. Together, the staged multi-agent architecture, skill-guided planning, and execution-informed replanning enable OceanMind to translate diverse requests into reliable workflows grounded in quantitative ocean evidence. Beyond reliable workflow generation, OceanMind provides a transparent and interactive analysis environment through a graphical user interface (GUI). The user can inspect analysis plans, follow the analysis execution, and examine intermediate results through the GUI. To summarize, OceanMind advances the integration of LLMs into time-dependent 3D ocean analysis, enabling researchers to focus on scientific interpretation and supporting non-specialists in formulating environmental policies grounded in quantitative ocean evidence.

To assess OceanMind's ability to analyze time-dependent 3D ocean state estimates, we evaluated OceanMind using three complementary perspectives. First, we evaluated whether OceanMind could translate diverse analytical requests into reliable computational workflows. For this first test, we constructed a benchmark of 240 queries spanning 59 target skills and four complexity levels (L1 to L4), using ocean state estimates from the China Sea Multiscale Ocean Modeling System (CMOMS)~\cite{gan2016numerical,gan2016three,lu2020nutrient}. The results comparison used ReAct, a widely used general-purpose agent paradigm that interleaves reasoning and tool use, with access to the same registered tool pool. OceanMind achieved a 91.2\% pass rate and 90.5\% effective step efficiency. Its performance remained robust across task complexity, with pass rates of 100\%, 96\%, 83\%, and 91\% from L1 to L4, respectively. Second, we applied OceanMind to diagnose regional oceans using CMOMS results. This demonstration resulted in OceanMind reproducing published ocean state diagnostics, constructing traceable evidence chains for hypothesis validation, and supporting environmental decisions using spatially explicit risk assessment. Finally, we evaluated whether OceanMind could generalize from a CMOMS-based regional diagnosis to a global ocean diagnosis using products from the Copernicus Marine Environment Monitoring Service (CMEMS)~\cite{le2019observation}. For the last test, we applied the same analytical framework to the Gulf of Mexico and tropical Pacific.

\section*{Results} \label{results}

\subsection*{OceanMind directly connects LLMs to time-dependent 3D multi-variables and physical-biogeochemical ocean states}

We developed OceanMind as a multi-agent system that directly analyzes time-dependent 3D ocean state estimates in response to natural-language requests. As shown in Fig.~\ref{fig:overview}a, OceanMind supports automated ocean diagnosis from regional to global scales, using CMOMS for the China Seas and global CMEMS products for the Gulf of Mexico region and tropical Pacific. The user submits natural language queries through the interactive analysis environment (that is, the GUI), and OceanMind returns synthesized responses, quantitative evidence, and corresponding visualizations (Fig.~\ref{fig:overview}b). Rather than presenting only a final answer, the interface exposes analysis plans, execution progress, and intermediate results, allowing the user to inspect how the evidence is produced so that they can interactively refine their analyses. Figure~\ref{fig:overview}c illustrates the architecture of OceanMind, which connects four layers: the user, GUI, analytical workflow, and underlying ocean state estimates. Unlike general-purpose LLMs that primarily retrieve relevant information from the Internet, OceanMind connects its analytical workflow directly to the data layer. This connection allows OceanMind to extract critical information through numerical computation and ground its responses in quantitative evidence.

OceanMind organizes the analytical workflow into four coordinated stages. {Query Routing} interprets each request and determines whether Internet information, active dataset computation, or both are required. {Skill-based Planning} translates the analytical request and its scientific constraints into a multi-step computational workflow. {Tool Execution} performs the planned computations on ocean state estimates, and {Evidence-based Summary Generation}  integrates the resulting artifacts into a response based on data. Six specialized LLM agents perform complementary functions across these stages to form OceanMind's collaborative multi-agent analytical architecture.

Ocean-analysis requests vary widely in their variables, regions, periods, depths, diagnostics, and expected outputs. Therefore, constructing a reliable computational workflow during {Skill-based Planning} is therefore particularly challenging; each request requires a different sequence of dependent operations. Complex mechanism ranking and evidence synthesis requests could involve more than 20 interdependent tool calls. To address this challenge, OceanMind incorporates 63 reusable ocean-specific analytical skills that are the procedural manuals for the PlannerAgent. Each skill specifies its scientific purpose, applicable and excluded scenarios, required evidence, optional inputs, compatible tools, and execution strategy. The PlannerAgent selects and composes relevant skills to resolve the required data scope and to organize the permitted operations into a dependent workflow. These skills guide the selection and coordination of 125 registered tools for data access, preprocessing, ocean diagnostics, event detection, statistical analysis, visualization, and evidence assembly. A deterministic ToolOrchestrator validates tool parameters, resolves artifact dependencies, executes the planned operations, and retains intermediate results for subsequent steps. With reflection and replanning mechanisms that use execution feedback to repair invalid plans, OceanMind ensures reliable analysis workflows across diverse analytical needs.

\begin{figure}[htp]
  \centering
  \includegraphics[width=0.99\textwidth,keepaspectratio]{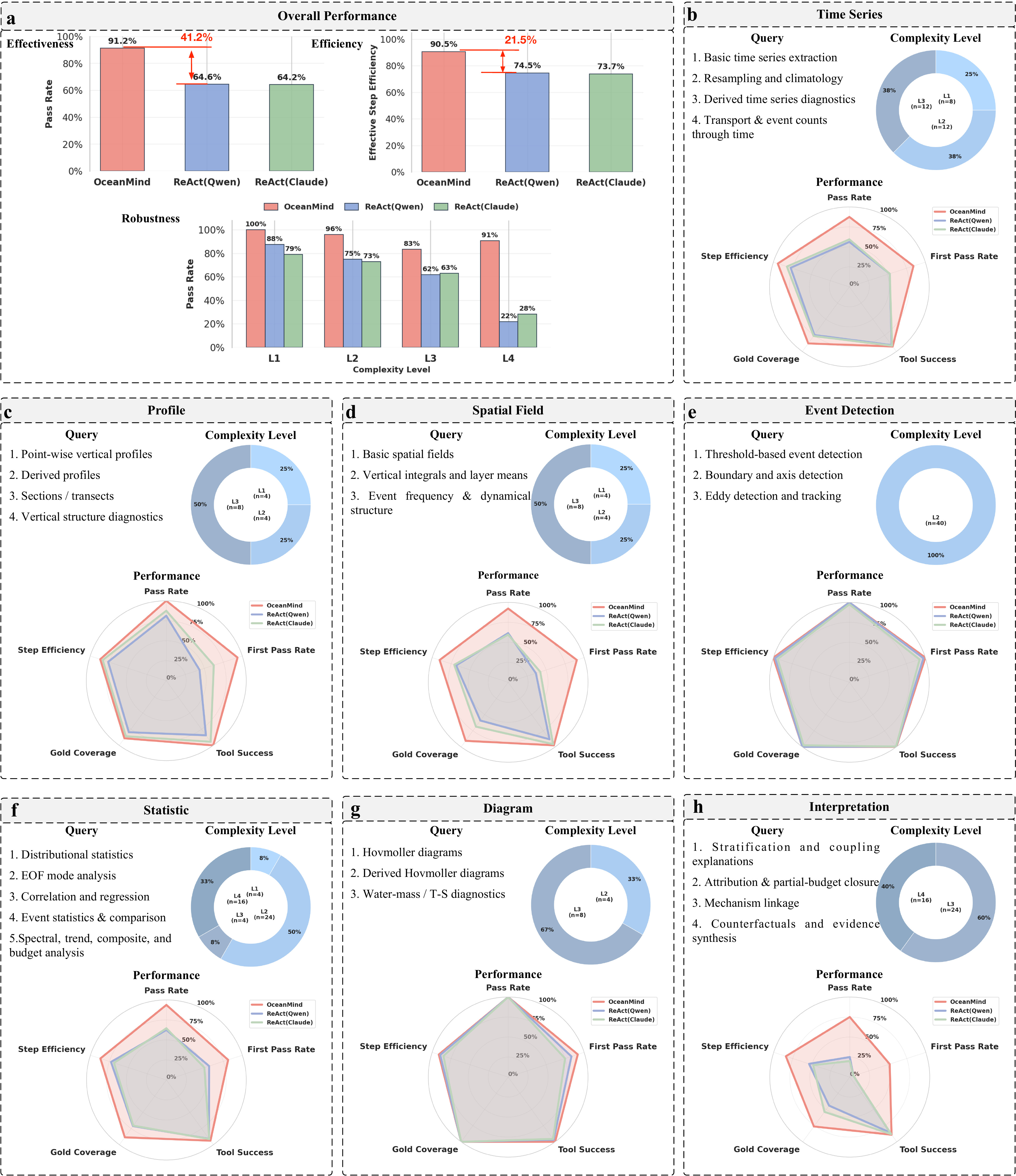}
  \caption{\textbf{Benchmarking OceanMind across ocean analytical tasks.}
\textbf{a.} Overall performance of OceanMind, ReAct(Qwen), and ReAct(Claude) across 240 queries. Effectiveness is the rule-reviewed pass rate. Efficiency is the mean effective step efficiency, defined as $\min(N_\mathrm{ref}/N_\mathrm{run},N_\mathrm{run}/N_\mathrm{ref})$ from the numbers of executed and reference core steps. Robustness is the pass rate stratified by L1 to L4 task complexity. Red annotations indicate OceanMind's relative improvement over the strongest ReAct baseline.
\textbf{b-h.} Benchmark composition and performance for time-series ($n=32$), profile ($n=16$), spatial-field ($n=52$), event-detection ($n=40$), statistical ($n=48$), diagram ($n=12$), and interpretation ($n=40$) queries. Each panel lists representative query families, shows the complexity-level distribution, and compares pass rate, first-pass rate, tool-success rate, gold-tool coverage, and effective step efficiency.
  }
  \label{fig:benchmarking}
\end{figure}

\subsection*{OceanMind outperforms competing AI agents on a benchmark of ocean analysis tasks spanning multiple complexity levels}


We designed the benchmark to evaluate whether an AI agent could turn inquiries about ocean states into executable and scientifically adequate analytical workflows for accurate and efficient solutions. We selected CMOMS as the benchmark data source because CMOMS provides high-resolution, time-dependent, 3D coupled physical-biogeochemical data that are rigorously validated and widely used in studies~\cite{gan2016numerical,gan2016three,liu2017pathways,gan2022hotspots}. The suite contained 240 natural language queries spanning the seven result types: time series, profile, spatial field, event detection, statistical diagnostics, specific diagram generation, and interpretation. Furthermore, the suite covered 59 oceanographic analysis skills and was stratified across four complexity levels: (1) \emph{L1} tasks required direct extraction or visualization; (2) \emph{L2} tasks combined preprocessing, event detection, resampling, or derived diagnostics; (3) \emph{L3} tasks required multi-step diagnostics such as Hovm\"oller diagrams, EOF analysis, correlations, event comparison and so on; and (4) \emph{L4} tasks required mechanism-oriented interpretation, budget reasoning, or evidence synthesis. Query variants further perturbed region, variable, time scale, visualization style, and analytical parameters, so success required both correct tool selection and faithful preservation of the user's scientific constraints.


We benchmarked OceanMind against two ReAct baselines~\cite{yao2023react} that shared the same ocean-tool registry (Fig.~\ref{fig:benchmarking}a). OceanMind and ReAct(Claude) both used Claude Sonnet 4.5. ReAct(Qwen) used Qwen3 Max. The ReAct baselines followed a standard reason-act-observe loop over the available tools, without OceanMind's staged routing, skill planning, execution validation, or reflection and re-planning steps. We evaluated each system using three criteria: (1) \emph{effectiveness}, whether the final answer was scientifically adequate and constraints were consistent; (2) \emph{efficiency}, how closely the number of executed core steps matched the canonical gold tool chain; and (3) \emph{robustness}, whether performance held when tasks demanded greater composition, constraint tracking, and evidence synthesis. OceanMind achieved a 91.2\% pass rate, compared to 64.6\% for ReAct(Qwen) and 64.2\% for ReAct(Claude), representing a 41.2\% relative improvement over the stronger ReAct baseline (Fig.~\ref{fig:benchmarking}a). OceanMind also reached 90.5\% effective step efficiency, compared to 74.5\% for ReAct(Qwen) and 73.7\% for ReAct(Claude), a relative improvement of 21.5\%. The complexity-stratified results showed that OceanMind remained reliable as workflow demands increased. Its pass rates were 100.0\%, 96.0\%, 83.3\%, and 90.6\% from L1 to L4, compared to 87.5\%, 75.0\%, 61.9\%, and 21.9\% for ReAct(Qwen), and 79.2\%, 73.0\%, 63.1\%, and 28.1\% for ReAct(Claude). The biggest differences occurred on L4 tasks, indicating that OceanMind was beter at organizing long, dependent analytical workflows while preserving the user's scientific constraints.

Fig.~\ref{fig:benchmarking}b-h further shows that differences across result types were governed by whether each category mainly contained obvious tool workflows or higher-level diagnostic composition. Beyond pass rate and first-pass rate, we reported three workflow diagnostics: tool-success rate measured the fraction of final-attempt tool calls that completed successfully; gold-tool coverage measured whether the required diagnostic operations were executed; and effective step efficiency measured agreement between the executed and reference core-step counts. For event-detection tasks (Fig.~\ref{fig:benchmarking}e), all 40 queries were \emph{L2}, and the three systems were nearly equal: OceanMind, ReAct(Qwen), and ReAct(Claude) reached pass rates of 100.0\%, 100.0\%, and 97.5\%, with first-pass rates of 100.0\%, 97.5\%, and 92.5\%, respectively. By contrast, result types that included \emph{L3} or \emph{L4} diagnostic-composition tasks showed clear overall gains, not just gains within the highly complexity subset.

Time-series queries (Fig.~\ref{fig:benchmarking}b), which spanned \emph{L1 to L3}, improved from a 53.1\% first-pass rate for ReAct(Qwen) to 84.4\% for OceanMind, and from a 56.2\% pass rate to 87.5\%. Spatial-field queries (Fig.~\ref{fig:benchmarking}d), which also contained \emph{L3} derived diagnostics, had the same pattern: OceanMind reached a 90.4\% first-pass rate and a 92.3\% pass rate, compared to 36.5\% and 61.5\% for ReAct(Qwen), respectively. The heavy statistical reasoning and interpretation categories (Fig.~\ref{fig:benchmarking}f, h) further amplified this effect. OceanMind reached a 93.8\% pass rate for statistical diagnostics and 75.0\% for interpretation, compared to, at most, 64.6\% and 25.0\%, respectively, for the ReAct baselines. OceanMind also matched or exceeded the baselines in gold-tool coverage and effective step efficiency, indicating that the planner reduced missing and redundant steps while preserving workflow completeness.

To assess whether the OceanMind framework depended on a specific LLM backend, we further compared OceanMind with Claude Sonnet 4.5, Qwen3 Max, and DeepSeek V4 Pro (Supplementary Fig.~S2). OceanMind still performed strongly across all three backends: Claude Sonnet 4.5 reached a 91.2\% pass rate and 90.5\% effective step efficiency; Qwen3 Max reached 88.8\% and 89.1\%; and DeepSeek V4 Pro reached 91.2\% and 89.4\%, respectively. More importantly, the advantage over the strongest ReAct baseline persisted on high-complexity tasks: the strongest ReAct baseline reached 63.1\% and 28.1\% pass rates on L3 and L4 tasks, whereas OceanMind with Claude Sonnet 4.5, Qwen3 Max, and DeepSeek V4 Pro reached 83.3\%/90.6\%, 85.7\%/75.0\%, and 84.5\%/84.4\%, respectively. These results indicated that OceanMind’s staged planning, validation, and re-planning workflow would remain effective even if the backend LLM were changed.

To summarize, the benchmark results showed that OceanMind improved the reliability of ocean analysis by structuring how open-ended scientific requests were routed, planned, executed, validated, and synthesized, rather than by expanding the tool set or relying on a single stronger LLM. The architecture produced its clearest gains when tasks required multi-step diagnostic composition, constraint tracking, or mechanism-oriented evidence synthesis, and, therefore, established OceanMind as a more reliable and efficient workflow system for complex ocean data analysis.

\begin{figure}[htp]
  \centering
  \includegraphics[width=0.99\textwidth,keepaspectratio]{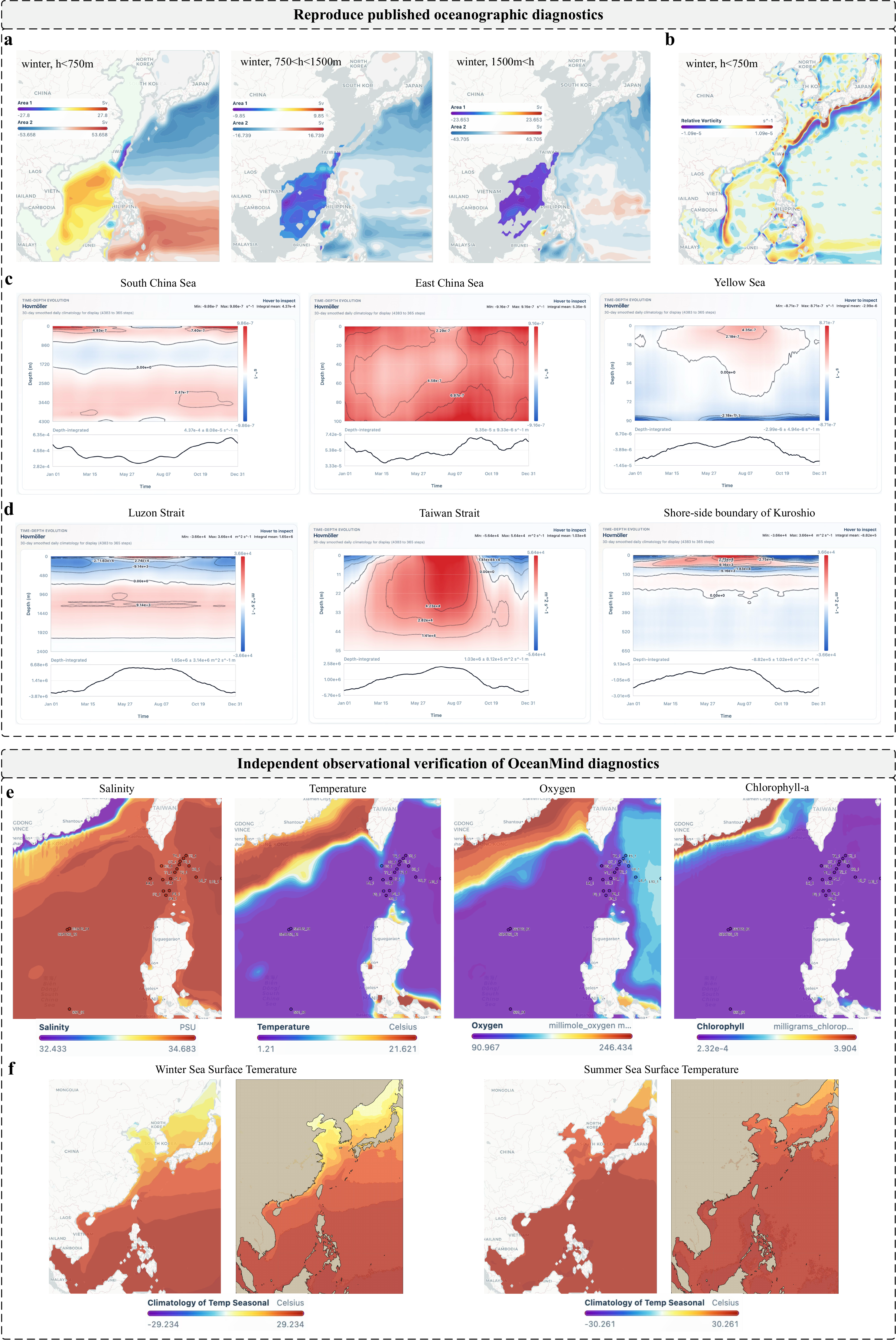}
\end{figure}
\begin{figure}[!t]
  \caption{
    \textbf{OceanMind reproduces published ocean diagnostics with independent observational verification.}
\textbf{a-d.} OceanMind reproduces selected kinematic diagnostics of China Seas circulation presented in Figs. 10, 11, 12, and 14 of Gan et al.~\cite{gan2016numerical}. These diagnostics characterize the intrinsic and extrinsic controls on three-dimensional circulation in the China Seas.
\textbf{a.} Winter transport stream functions show a strong upper-layer field ($h<750$ m), a much weaker and predominantly negative middle-layer field ($750<h<1500$ m), and a deep field ($h>1500$ m) with large negative and positive extrema.
\textbf{b.} Winter relative vorticity in the upper 750 m has a domain mean close to zero but closely spaced positive and negative extrema near the Philippine Sea, indicating localized mesoscale activity.
\textbf{c.} Time-depth diagrams and depth-integrated time series show weak but variable area-averaged relative vorticity in the South China Sea, East China Sea, and Yellow Sea, with the strongest signals generally near the surface and intermittent excursions at depth.
\textbf{d.} Time-depth normal flux and depth-integrated transport show strong interannual variability and upper-ocean-intensified extremes at the Luzon Strait, Taiwan Strait, and the Kuroshio's shore-side boundary.
\textbf{e-f.} Independent verification of OceanMind diagnostics using in situ and remote-sensing observations.
\textbf{e.} Nineteen bottom CTD observations from the \texttt{201801\_XMU\_CHOICE-II} downcast profiles overlay January 2018 bottom salinity, temperature, oxygen, and chlorophyll-a fields generated by OceanMind in the South China Sea.
\textbf{f.} Paired maps compare winter and summer sea surface temperature climatologies generated by OceanMind from CMOMS during from 2011 to 2021 with the independent GHRSST MUR Level 4 SST analysis.
Continuous-map colorbars use robust display ranges rather than the full-field extrema reported in Supplementary Sections~3.1-3.6.}
  \label{fig:3}
\end{figure}

\subsection*{OceanMind reproduces published ocean diagnostics with independent observational verification}

The benchmark results establish that OceanMind can construct reliable workflows for ocean data analysis. Next, we evaluated whether this capability can accelerate scientific discovery by reproducing a published dynamic analysis of the ocean. By directly extracting quantitative information from time-dependent 3D ocean state estimates, OceanMind enables the user to generate and inspect literature-level oceanographic diagnostics by interacting with the GUI using natural language. In contrast, existing systems cannot support the same end-to-end diagnosis: general LLMs like ChatGPT are not directly connected to active ocean state estimates, and OceanAI lacks access to comprehensive time-dependent 3D ocean state estimates and the analytical functions required by such a multi-step workflow.

We selected the China Seas circulation study of Gan et al.~\cite{gan2016numerical,noaaOisst} the authors identified vertically alternating rotating circulation through a connected sequence of transport stream function, Stokes' theorem, and vorticity dynamics. This case, therefore, evaluated whether OceanMind could reconstruct a multi-step physical diagnosis rather than generate isolated plots. We applied OceanMind to daily velocity output of CMOMS from 2011 to 2022 to reproduce the China Seas circulation diagnostics presented in Figs. 10, 11, 12, and 14 of Gan et al.~\cite{gan2016numerical} (here, Fig.~\ref{fig:3}a-d). The complete reproduction queries, computational workflows, intermediate results, and evidence-based summaries are provided in Supplementary Sections~3.1 to 3.4. In the winter fields shown in Fig.~\ref{fig:3}a, the upper-layer stream function ranged from $-53.7$ to $+38.7$ Sv, compared to $-16.7$ to $+4.3$ Sv in the weaker, predominantly negative middle-layer field and $-43.7$ to $+22.1$ Sv in the deep field. Figure~\ref{fig:3}b shows a winter upper-layer relative-vorticity field with a domain mean of $7.38\times10^{-8}$ s$^{-1}$ and localized extrema of $+6.77\times10^{-5}$ and $-8.75\times10^{-5}$ s$^{-1}$ near the Philippine Sea. Figure~\ref{fig:3}c presents 30-day-smoothed time-depth climatologies of area-averaged relative vorticity computed from 4,383 daily fields. The South China Sea and East China Sea fields had weak but variable signals, with the strongest positive anomalies near the surface and intermittent negative excursions at depth. The Yellow Sea showed weak overall vorticity with intermittent bursts, strongest near the surface and occasionally extending deeper. Figure~\ref{fig:3}d examines three representative exchange pathways. The full-depth Luzon Strait mean transport was 1.245 Sv and ranged from $-14.95$ to $+17.46$ Sv, with extreme fluxes concentrated in the upper 1000 m. The Taiwan Strait mean transport over 0 to 60 m was $+1.00$ Sv but underwent frequent reversals, whereas the mean transport across the 0 to 700 m Kuroshio shore-side transect was $-0.759$ Sv and ranged from $-8.601$ to $+4.253$ Sv. Together, these diagnostics connect the spatial circulation fields to the temporal variability of major exchange pathways without inferring extrema from the plotted colorbar limits.

We further evaluated the generated ocean-state fields using independent in situ and remote-sensing observations (Fig.~\ref{fig:3}e, f). Figure~\ref{fig:3}e overlays 19 bottom CTD observations from the \texttt{201801\_XMU\_CHOICE-II} downcast profiles on January 2018 and bottom salinity, temperature, oxygen, and chlorophyll-a fields generated by OceanMind. The observations followed the major spatial gradients represented in the generated fields. Figure~\ref{fig:3}f compares winter and summer SST climatologies generated by OceanMind from CMOMS during from 2011 to 2021 with independent GHRSST MUR Level 4 SST fields. Both products reproduced the large-scale winter temperature gradient across the China Seas and the widespread surface warming during summer. Detailed queries and outputs for Fig.~\ref{fig:3}e and Fig.~\ref{fig:3}f are provided in Supplementary Sections~3.5 and 3.6, respectively.

Together, the published diagnostic reproduction and independent observational comparisons demonstrated that OceanMind reconstructed connected analyses from time-dependent 3D ocean state estimates. The system exposed each computational step and its resulting evidence through the GUI, allowing the user to inspect the complete diagnostic process rather than receive an isolated visualization or unsupported textual conclusion.

\begin{figure}[!t]
  \centering
  \includegraphics[width=0.99\textwidth, keepaspectratio]{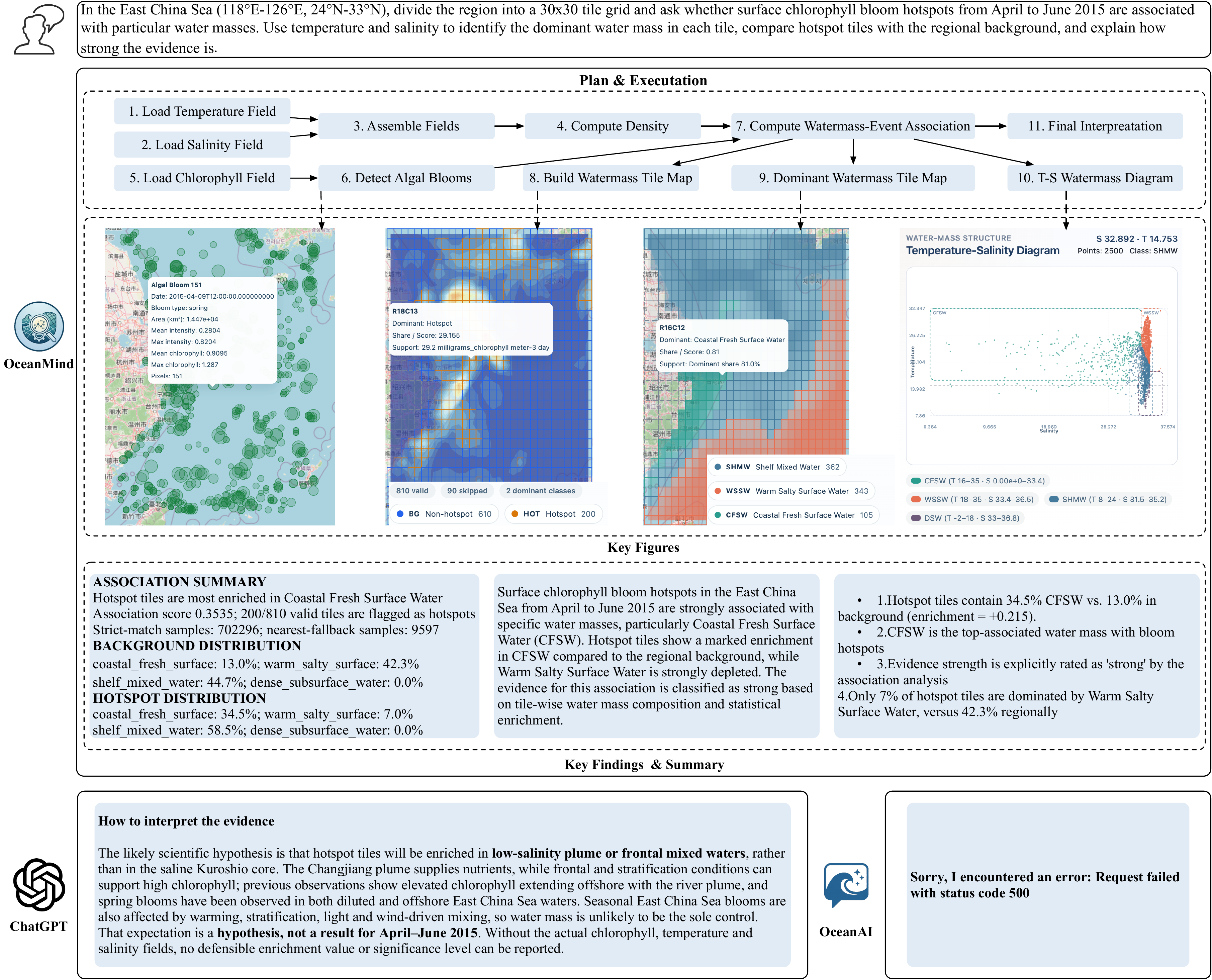}
  \caption{\textbf{Comparing how well OceanMind, ChatGPT, and OceanAI validate a scientific hypothesis.}
  Each system was asked whether there were chlorophyll-a bloom hotspots in the East China Sea from April to June 2015 associated with particular water masses. OceanMind executed a traceable 11-step workflow and identified enriching Coastal Fresh Surface Water in hotspot tiles (34.5\% versus 13.0\% in the regional background). ChatGPT provided a plausible literature-based explanation without a quantitative analysis, whereas OceanAI did not complete the request in this run.
  }
  \label{fig:hypothesis-validation}
\end{figure}

\subsection*{OceanMind enables traceable validation of scientific hypotheses}

Scientific hypotheses that govern oceanic processes often ask whether one process is more plausible than another, not merely whether a field can be mapped. Answering such questions requires an evidence chain that links multiple variables and complex processes through data transformation, diagnostics, uncertainty quantification, and interpretation. OceanMind supports this mode of research by converting a user-stated hypothesis into an executable diagnostic workflow and exposing the intermediate evidence on which acceptance, rejection, or qualification of the hypothesis rests. By contrast, general-purpose LLMs, such as ChatGPT, can use internet searches to provide seemingly relevant but qualitative information without deriving quantitative results from active ocean state estimates. OceanAI, meanwhile, cannot analytically execute such complex multi-step hypothesis-validation workflows.

Figure~\ref{fig:hypothesis-validation} examines the hypothesis that surface chlorophyll-a bloom hotspots occurred in the East China Sea from April to June 2015 and were associated with particular water masses. OceanMind evaluated this hypothesis using an 11-step workflow integrating temperature, salinity, and chlorophyll-a fields, bloom detection, water-mass classification, and comparing hotspots with their backgrounds. Of 810 valid tiles, OceanMind identified 200 tiles as bloom hotspots. Areas with Coastal Fresh Surface Water accounted for 34.5\% of hotspot tiles compared to 13.0\% of the regional background, representing an approximately 2.7-fold enrichment. Conversely, areas with Warm Salty Surface Water accounted for only 7.0\% of hotspot tiles compared to 42.3\% of the background. These contrasting proportions supported the hypothesis by demonstrating that bloom hotspots preferentially coincided with areas with Coastal Fresh Surface Water but did not establish a causal relationship. The computational plan and intermediate bloom, water-mass, and temperature–salinity diagnostics remained available for the user to inspect so that the evidence underlying the assessment was traceable. In comparison, ChatGPT provided a plausible literature-based explanation linking blooms to low salinity Changjiang River plume water and frontal mixing. However, ChatGPT did not analyze the underlying fields and provided no quantitative evidence for the specified period. OceanAI returned a server error in this run and provided no computed evidence nor an analytical conclusion.

\subsection*{OceanMind supports evidence-based policy making}
\begin{figure}[htp]
  \centering
  \includegraphics[width=0.99\textwidth,keepaspectratio]{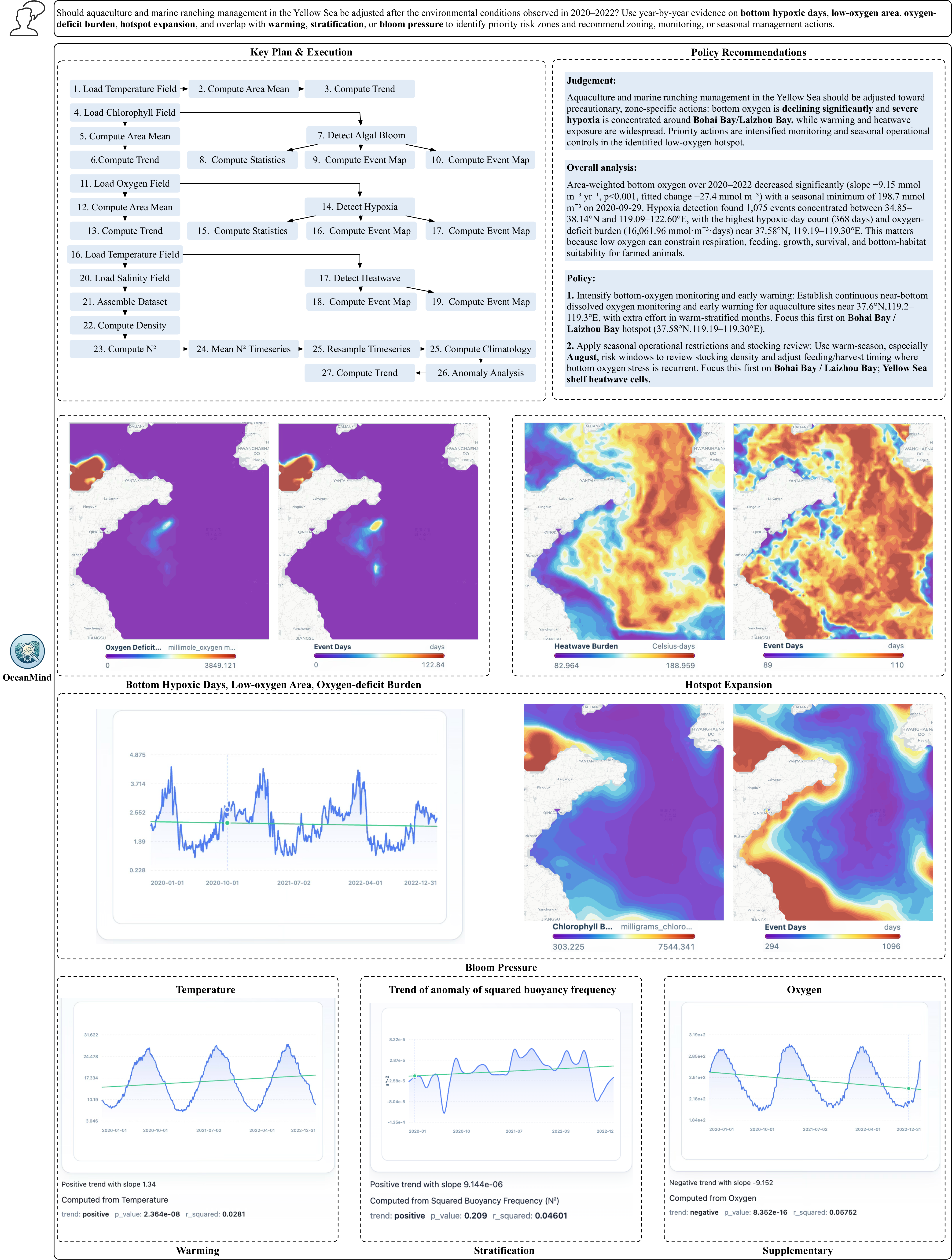}
\end{figure}
\begin{figure}[!t]
  \caption{\textbf{OceanMind supports evidence-based policy making.}
OceanMind assesses whether aquaculture and marine ranching management in the Yellow Sea should be adjusted in response to environmental conditions observed during 2020-2022.
The workflow integrates bottom-oxygen trends, hypoxia timing and burden, heatwave and bloom exposure, and squared buoyancy frequency (\(N^2\)) to identify priority regions and generate evidence-bounded management recommendations.
The displayed recommendations prioritize enhanced bottom-oxygen monitoring at the identified low-oxygen hotspot and seasonal adjustments to stocking, feeding, and harvest timing.
The trend in \(N^2\) anomalies is not statistically significant, so stratification provides seasonal timing context rather than evidence of progressively strengthening stratification.
Continuous-map colorbars show robust display ranges; the hotspot values reported in the policy summary are calculated from the underlying full fields and can exceed these displayed limits.
  }
  \label{fig:marine-management}
\end{figure}

OceanMind extends time-dependent 3D ocean diagnosis to evidence-based policy making. To demonstrate this functionality, we constructed complementary policy-oriented skills that connected management objectives to ocean diagnostics. Using the example of deoxygenation that is threatening the global ocean~\cite{breitburg2018declining}, we demonstrated that OceanMind could provide the environmental health assessment by assembling evidence of bottom oxygen stress, hypoxia, warming, stratification, and bloom pressure. The policy recommendation skill then translated the results into evidence-bounded actions while distinguishing quantitative evidence, indirect reasoning, and data gaps.

Figure~\ref{fig:marine-management} shows how OceanMind assessed whether the Yellow Sea aquaculture and marine ranching management should change in response to environmental conditions from 2020 to 2022. The workflow integrated temperature, salinity, oxygen, and chlorophyll-a fields to characterize environmental trends, hypoxia, marine heatwaves, bloom exposure, and stratification. Area-weighted bottom oxygen declined significantly over the study period  (slope \(=-9.15\) mmol m\(^{-3}\) yr\(^{-1}\), \(p<0.001\)). The computed fields identified a low-oxygen hotspot near \(37.58^\circ\)N, \(119.19^\circ\)E to \(119.30^\circ\)E, with up to 368 hypoxic days and an oxygen-deficit burden of 16,062 mmol m\(^{-3}\) days. Warming and marine heatwave exposure provided additional environmental-risk context~\cite{oliver2018longer,frolicher2018marineheatwaves,smale2019marine}, with heatwave exposure reaching 110 days at individual grid cells. The squared buoyancy frequency (\(N^2\)) time series showed seasonal variability, but the anomaly trend was not statistically significant (\(p=0.209\)). Stratification, therefore, provided seasonal context rather than evidence of progressively strengthening stratification.

These results supported precautionary, spatially targeted management rather than unconditional changes to aquaculture zoning. OceanMind translated the assessment into two recommendations. First, it recommended continuous near-bottom oxygen monitoring and early warning around the identified low oxygen hotspot, with intensified monitoring during warm, stratified months. Second, it recommended reviewing stocking density and adjusting feeding and harvest timing during warm season risk windows, particularly in August. The recommendations prioritized the low oxygen hotspot and, for seasonal operational measures, areas of the Yellow Sea shelf that are exposed to heatwaves. Together, these actions connected the computed environmental evidence to specific monitoring priorities and seasonal management decisions, without treating the insignificant stratification trend as evidence of increasing risk.

In contrast, ChatGPT outlined the assessment, but without direct data access, ChatGPT had to assemble internet information and previously published analyses. Likewise, OceanAI could not complete the request because OceanAI's supported products and functions excluded vertically resolved, coupled physical-biogeochemical diagnostics. Our supplementary comparison clarified the difference between computed and literature-assembled evidence (Supplementary Figs. S3 and S4). For the 2020 to 2022 period, ChatGPT retrieved separate studies on blooms and hypoxia. Then, ChatGPT acknowledged the absence of comparable Yellow Sea indicators yet assumed that the missing indicators were consistently calculated before presenting its management judgment. ChatGPT's response, therefore, combined previously published findings rather than performing a complete analysis of the study period. For the 2030 to 2032 future period, ChatGPT found no observed evidence and could only describe how a forecast-based assessment might be conducted. If projected time-dependent 3D ocean fields were assembled in OceanMind, the same workflow could compute these indicators and formulate recommendations grounded in the projections.

\begin{figure}[!t]
  \centering
  \includegraphics[width=0.99\textwidth,height=0.68\textheight,keepaspectratio]{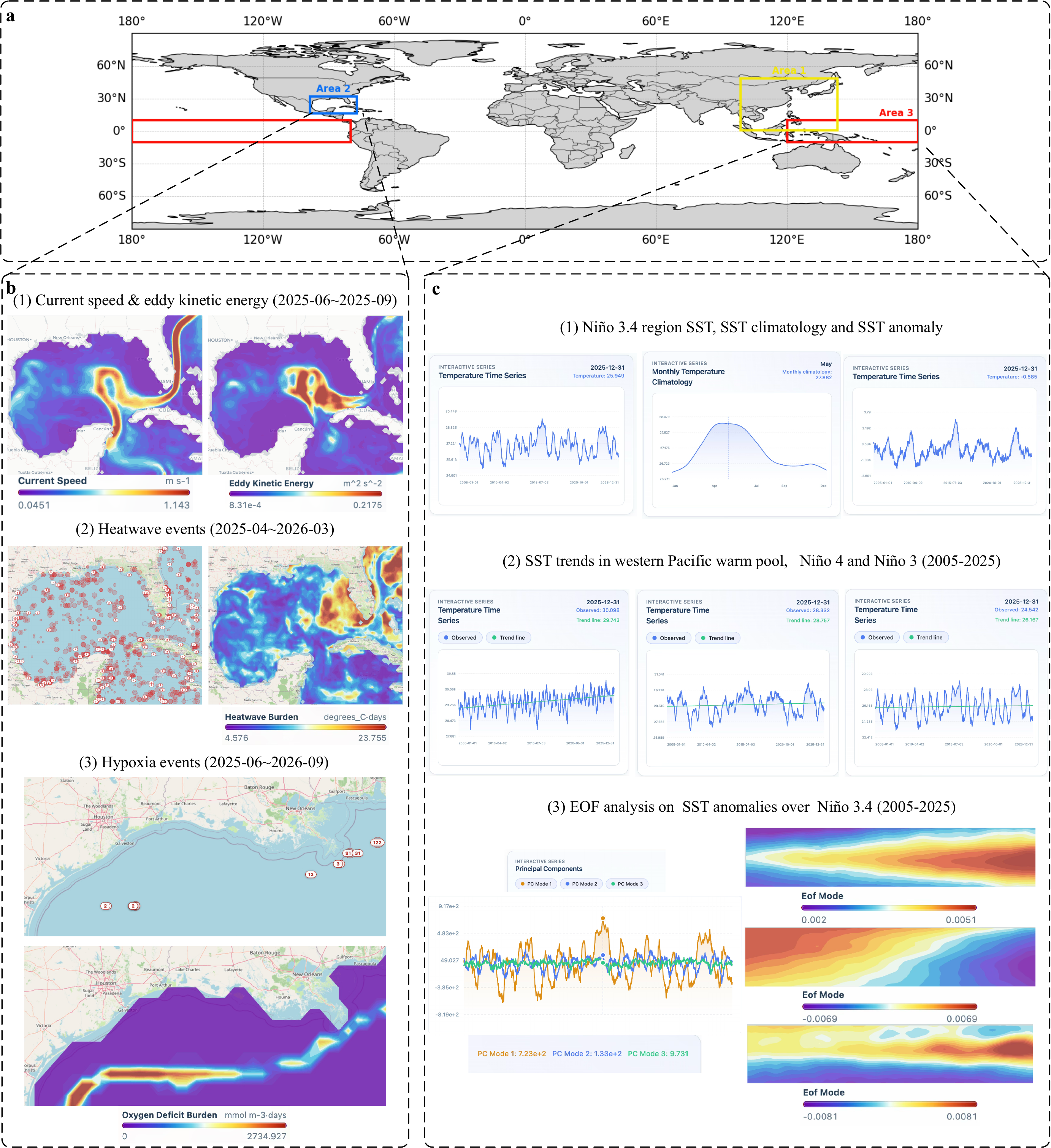}
  \caption{\textbf{OceanMind can be applied globally.}
\textbf{a.} Global map showing the representative CMEMS analysis domains: Area 2, from 98.68$^\circ$W to 76.98$^\circ$W and from 16.35$^\circ$N to 32.02$^\circ$N, and Area 3, from 120$^\circ$E to 80$^\circ$W and from 10$^\circ$S to 10$^\circ$N.
\textbf{b.} Area 2 diagnostics from full-depth CMEMS products in the Gulf of Mexico: June to September 2025 mean surface current speed and eddy kinetic energy; marine heatwave event timing, extent, intensity, cumulative exposure, and thermal burden for the analysis period from 1 April 2025 to 31 March 2026; and relatively low bottom-oxygen exposure from June to September 2025, identified using a relaxed threshold of 150 mmol m$^{-3}$.
\textbf{c.} Area 3 tropical Pacific diagnostics from the CMEMS surface layer (0.494 m) from 1 January 2005 to 31 December 2025, including EI Ni\~no 3.4 SST, climatology and anomaly, regional SST trends, and EOF modes of EI Ni\~no 3.4 SST anomalies.
For continuous maps, the graphical user interface used robust 2nd-98th percentile display limits (or the symmetric 98th percentile of absolute values for fields crossing zero). Reported extrema were calculated from the full fields and might exceed the displayed colorbar limits.
  }
  \label{fig:global_generalization}
\end{figure}

\subsection*{OceanMind can generalize across ocean products and regions}

Fig.~\ref{fig:global_generalization} demonstrates that OceanMind could be applied to global ocean analysis using CMEMS\footnote{\url{https://marine.copernicus.eu/}} products as the common data source. CMEMS combines satellite observations, \emph{in-situ} measurements and numerical ocean model results to provide global and regional products for physical and biogeochemical variables, and has been widely used in ocean research, operational monitoring, forecasting and marine environmental assessment~\cite{le2019observation}. This wide variety of data makes CMEMS a suitable testbed for evaluating whether OceanMind can generalize beyond a single regional ocean state estimate. We focused on two representative domains outside the China Seas: Area 2 covered the Gulf of Mexico and northern Caribbean, from 98.68\(^{\circ}\)W to 76.98\(^{\circ}\)W and from 16.35\(^{\circ}\)N to 32.02\(^{\circ}\)N. Area 3 covered the tropical Pacific, from 120\(^{\circ}\)E to 80\(^{\circ}\)W and from 10\(^{\circ}\)S to 10\(^{\circ}\)N. These examples assessed whether OceanMind could use the same natural language workflow for diagnostics in different oceanic settings, from regional coastal hazards to basin-scale climate variability.

For Area 2, OceanMind diagnosed conditions in the Gulf of Mexico using physical and biogeochemical CMEMS products. We replaced the CMOMS in Fig.~\ref{fig:overview}c. OceanMind leveraged Data, Field, and Event Detection Skill in its analysis workflow with the CMEMS products. The mean fields from June to September 2025 had fast boundary current flow near Florida Strait and elevated eddy activity in the central Gulf of Mexico (Fig.~\ref{fig:global_generalization}b)~\cite{piecuch2020florida,lin2017lagrangian}. Mean surface current speed was 0.30 m s\(^{-1}\) and reached a full-field maximum of 1.79 m s\(^{-1}\) near 27.25\(^{\circ}\)N, 79.75\(^{\circ}\)W. Eddy kinetic energy had a mean of 0.036 m\(^{2}\) s\(^{-2}\), a median of 0.010 m\(^{2}\) s\(^{-2}\), and a full-field maximum of 0.37 m\(^{2}\) s\(^{-2}\) near 25\(^{\circ}\)N, 88\(^{\circ}\)W. One should interpret these patterns cautiously because the source dataset contained inconsistent metadata. For the marine heatwave analysis from 1 April 2025 to 31 March 2026, OceanMind detected 1,251 events occurring from 23 May to 28 November 2025. The largest event covered 1,868,949 km\(^{2}\) on 22 August. The strongest heatwave reached 1.92\(^{\circ}\)C above threshold on 30 August. Cumulative exposure reached 37 days, and thermal burden exceeded 41\(^{\circ}\)C-days near the Florida Straits. A strict hypoxia threshold detected no events. Using the requested relaxed threshold of 150 mmol m\(^{-3}\), OceanMind, instead, identified 415 hypoxia-like events from June to September 2025. Relatively low bottom oxygen persisted for all 122 days at the most persistent location near 27.0\(^{\circ}\)N, 96.25\(^{\circ}\)W. Oxygen-deficit burden peaked at 3215 mmol m\(^{-3}\) days near 27.75\(^{\circ}\)N, 94.5\(^{\circ}\)W, where the largest event covered 32,127 km\(^{2}\) on 30 June. These analyses demonstrate how the same workflow framework was able to characterize circulation, thermal extremes, and low-oxygen exposure while retaining the thresholds and quantitative evidence used for each diagnosis.

For Area 3, OceanMind performed tropical Pacific climate diagnostics using the CMEMS surface temperature from 1 January 2005 to 31 December 2025. OceanMind utilized Statistical Skill in Fig.~\ref{fig:overview}c in the analysis workflow. In Fig.~\ref{fig:global_generalization}c(1), the EI Ni\~no 3.4 anomaly reached a record maximum of $+3.09^{\circ}$C on 1 December 2015, and raw SST reached 29.74$^{\circ}$C on 21 November 2015. The 95th-percentile anomaly threshold was $+1.73^{\circ}$C. Figure~\ref{fig:global_generalization}c(2) shows positive SST trends of $+0.0305^{\circ}$C yr$^{-1}$ in the western Pacific warm pool~\cite{weller2016warmpool}, $+0.0132^{\circ}$C yr$^{-1}$ in EI Ni\~no 4, and $+0.0112^{\circ}$C yr$^{-1}$ in EI Ni\~no 3. The warm-pool trend had the strongest explanatory power ($R^{2}=0.201$), whereas the two Ni\~no-region trends had $R^{2}<0.02$ and strong ENSO-related interannual variability. In Fig.~\ref{fig:global_generalization}c(3), the first EOF mode explained 82.6\% of the EI Ni\~no 3.4 anomaly variance, and the principal component peaked during the 2015 to 2016 EI Ni\~no~\cite{chen2017superelnino}. The second mode explained 9.9\% and exhibited a west-east dipole. Together, the Area 2 and Area 3 examples showed that OceanMind could be applied across ocean products and regions while preserving a transparent sequence from CMEMS data access to derived diagnostics and final interpretation. All results could be obtained through the OceanMind GUI (Fig.~\ref{fig:overview}c).

\section*{Discussion and summary}

In this study, we presented OceanMind, a conversational multi-agent AI system that directly couples LLMs with time-dependent 3D multi-variables and physical-biogeochemical states of the ocean to diagnose, validate scientific hypotheses, and provide support for evidence-based policy making. OceanMind automates analyses that are organized as four coordinated stages: Query Routing, Skill-based Planning, Tool Execution, and Evidence-based Summary Generation. Across these stages, specialized agents interpret the user's intent, construct and execute computational workflows, and synthesize quantitative evidence into transparent responses. The GUI exposes analysis plans, execution progress, and intermediate results, allowing the user to inspect the analysis as it develops. Together, these components connect natural language requests to reproducible maps, time series, diagnostics, evidence tables, and scientific interpretations. With these unique features, OceanMind outperforms competing AI agents on a benchmark of ocean analysis and is an ocean AI agent that has novel and comprehensive capabilities.

Among the four stages, Skill-based Planning is central because diverse analytical requests require different sequences of interdependent computations. Therefore, OceanMind uses 63 reusable ocean-specific analysis skills as procedural manuals for the PlannerAgent. These skills guide the selection of variables, regions, periods, depths, diagnostics, and appropriate tools from a registry of 125 members. A deterministic ToolOrchestrator executes the resulting workflow, while reflection and replanning return execution failures or unresolved dependencies to the PlannerAgent for correction. This skill-guided multi-agent design defines OceanMind's advantage over ReAct agents using the same tool pool, particularly on L3 and L4 tasks requiring complex workflow composition.

The results from our test runs also show that OceanMind is much more than a plotting interface. In the reproduction case, OceanMind reconstructed literature level circulation diagnostics from validated ocean datasets and their inherent physics. Independent checks against in situ and satellite remote sensing observations further supported the correctness of the generated diagnostics. In the hypothesis validation cases, OceanMind exposed the evidence chain behind each conclusion, including uncertainty and residual terms when support was incomplete. In the management case, OceanMind connected ocean hazards such as hypoxia, warming, stratification, bloom pressure, and trend diagnostics to spatially explicit recommendations. The test results showed how OceanMind could support scientific research, hypothesis testing and evidence-based marine management without hiding the computations behind a free-form text answer. Moreover, we designed OceanMind to be extensible across ocean state estimates. The global applications showed that the same workflow could be transferred to other regions after data were converted into compatible variables, coordinates, time ranges, and depth dimensions.

Despite all the advantages, OceanMind does have limits. First, the quality of OceanMind's conclusions is bounded by the quality and coverage of the underlying ocean state estimates. Biases in temperature, salinity, currents, oxygen, or chlorophyll fields can propagate into derived diagnostics and the final synthesis. Second, LLM agents can still misinterpret ambiguous questions or overstate weak evidence. OceanMind reduces this risk through routing, validation, reflection, and evidence-bounded synthesis, but these safeguards do not replace expert review~\cite{messeri2024illusions}. Third, OceanMind only supports vital physical and biogeochemical variables, but many questions require other variables or immediate products from ocean state estimates.

Therefore, future development should focus on stronger uncertainty handling, broader data coverage, and tighter coupling with numerical ocean modelling. Ensemble state estimates, satellite products, and field observations could be integrated to quantify uncertainty and distinguish robust signals from product specific artifacts. Additional skills could cover biogeochemical budgets, ecosystem indicators, source attribution, forecast verification, and data assimilation diagnostics. In conclusion, OceanMind advances ocean AI from text-based assistance toward executable, traceable, and reusable analysis of ocean state estimates. It provides a foundation for future ocean AI measurement and modelling systems that can support scientific discovery and environmental decision-making at regional and global scales.

\section*{Methods}

\subsection*{Architecture of OceanMind}

OceanMind converts natural language ocean questions into executable and reproducible diagnostic workflows through four stages (Fig.~\ref{fig:overview}c): (1) Query Routing determines whether the active dataset is needed or whether the request can be answered directly; (2) Skill-based planning selects the relevant ocean skills and turns the user's intent into a structured tool plan; (3) Tool execution runs the plan over gridded ocean state data, validates intermediate results, and revises the plan when needed; and (4) Evidence-based summary generation integrates the completed steps, computes diagnostics and visual evidence into a scientific answer grounded in the computed evidence.

To stabilize the full workflow, OceanMind used a multi-agent architecture with six LLM agents and one deterministic ToolOrchestrator. Query Routing first required a semantic decision about whether the user request needed active ocean state data. OceanMind therefore used a QueryProcessor to classify the request as dataset analysis or general answer. Requests outside the active dataset were passed to an AnswerAgent, which returned a direct response or requested external search when current information was needed. Dataset analysis requests entered Skill-based planning, where the PlannerAgent matched the question to ocean skills, dataset constraints and tool contracts, then composed an executable workflow. Tool execution used the ToolOrchestrator to resolve artifact references, call registered Python tools, normalize outputs and cache intermediate results. When no registered tool covered a required diagnostic operation, the CodingAgent generated a restricted Python step from declared inputs, expected outputs and the intended analysis. Evidence-based summary generation used the ResultSynthesizer to integrate completed computations, limitations and visual evidence. For multi-turn use, the MemoryAgent selected role-specific memory before planning or synthesis and recorded reusable definitions, preferences, artifacts and workflow patterns after each turn without overriding explicit user constraints.

In the OceanMind configuration we used in this study, the skill pool contains 63 reusable ocean skills and the tool pool contains 125 registered tools for ocean analysis. Each skill describes a class of ocean analysis tasks, including applicable scenarios, required and optional inputs, parameters that could be inferred from the query, permitted tools, and execution strategy. During planning, the selected skills guide the PlannerAgent to extract variables, regions, time ranges, depth selections, and analysis options from the user query and to organize the permitted tools into an executable workflow. The tool pool supports data access, diagnostics, event detection, visualization, and evidence assembly.

Moreover, we equipped OceanMind with a GUI to support interactive use. The user can submit queries, inspect selected plans, follow execution progress, and examine intermediate maps, charts, and result cards before reading the final synthesis. While developing OceanMind, we found that ocean state estimates were large and that direct computation could be inefficient for long analysis workflows. OceanMind, therefore, keeps loaded subsets as lazy partitioned arrays whenever possible. Partition-aware tools submit compatible partitions to a thread worker pool, computes the partitions in parallel, and merges the outputs into one normalized result. Asynchronous generators stream partition progress and step completion events to the GUI while computation continues.

\subsection*{Query Routing}

The Query Routing stage first determines whether the active ocean state estimate is required (Supplementary Section~1.2). It receives the user query, dataset context, structured inputs from the interface and recent conversation context, and returns a structured routing decision. Requests that require dataset computation, dataset inspection, or active dataset metadata are sent to the dataset analysis pathway. General ocean science questions, text editing requests, and other requests that do not require the active dataset are passed to the AnswerAgent, which produces a direct answer or requested external search for current information.

\subsection*{Skill-based planning}

OceanMind maintains 63 skills under the skills/directory. Each skill is a SKILL.md document that describes its scientific purpose, applicable and excluded scenarios, required evidence, optional inputs, compatible tools, and expected outputs. The skill library covers data loading, spatial fields, time series, profiles, sections, Hovmöller diagrams, climatology, anomaly and trend analysis, event detection, event statistics, vertical structure, transport, water masses, EOF analysis, budget diagnostics, mechanism ranking, evidence synthesis, and environmental health assessment for policy support.

Skill-based planning has three tasks (Supplementary Section~1.3). First, the PlannerAgent resolves the data scope needed by the request, including variables, region, time window, season, and vertical selection. Second, the PlannerAgent selects one skill or a minimal skill set that can generate the required evidence. Finally, the PlannerAgent organizes the selected operations into a traceable evidence workflow whose steps produce named intermediate artifacts and one or more final artifacts for interpretation.

Skill-based planning is constrained by selected skill workflows, dataset metadata, and registered tool contracts, so each step has known inputs, outputs, and artifact dependencies. If a required input cannot be inferred from the query, interface state, dataset metadata, or conversation context, the PlannerAgent returns a clarification request instead of inventing a parameter.

\subsection*{Tool execution}

The Tool execution stage runs the planned workflow over the ocean state estimates (Supplementary Section~1.4). Before a registered-tool step is executed, the tool name, arguments, and references are checked against the tool contract. Scalar and array parameters are normalized where possible, and invalid parameters are reported before the tool is called. These checks prevent unsupported arguments or broken references from being silently propagated through the workflow.es from being silently propagated through the workflow.

The ToolOrchestrator resolves all \texttt{\$ref:} dependencies, calls the target Python tool, normalizes the returned object, and caches the result under a stable identifier. Loaded ocean fields and derived datasets are wrapped as data artifacts with metadata. The ToolOrchestrator summarizes structured outputs, including event maps, trend diagnostics, section results, mechanism scores, and assessment records into compact records for the interface and for synthesis. The full numerical objects remain available to downstream tools.

When the required analysis is not covered by the registered ocean tool pool, OceanMind invokes a CodingAgent. The CodingAgent uses the required input artifacts, input schemas, expected output type, and intended diagnostic operation to design a restricted Python analysis step. The generated code is required to define a \texttt{run(inputs, params)} function, to use only approved scientific libraries, and to return a normalized result matching the expected output contract. Then, the execution layer runs the generated step, validates its output, and stores the output as a standard artifact.

This execution design allows workflows with multiple steps to pass evidence forward as structured artifacts. For example, a workflow could load bottom oxygen, detect hypoxia, compute an oxygen deficit burden map, extract regional statistics, and pass the resulting artifacts into an environmental assessment. Similarly, a dynamic workflow could load currents, compute vorticity or transport diagnostics, and use the resulting fields or time series in mechanism ranking or evidence synthesis.

\subsection*{Evidence-based summary generation}

After Tool execution, OceanMind collects result summaries, step cards, result cards, workspace data prepared for visualization, and compact evidence packets (Supplementary Section~1.5). The evidence packets avoid full numerical arrays and expose only the quantities needed for interpretation, such as spatial extrema, trend slopes, event counts, burden values, mechanism scores, and so on. having these packets keeps the synthesis grounded in completed computations while limiting the raw data passed to the language model.

The ResultSynthesizer converts these structured execution products into a scientific answer. It receives the original query, the workflow, completed step records, result summaries, and optional evidence packets. The synthesizer's output is validated against a schema and can include a main summary, findings, limitations, recommended follow-up analyses, policy guidance, assessment threads, evidence tables, and source cards. For integrated environmental health and suitability questions, a dedicated assessment profile distinguishes computed evidence, indirect support, and data gaps.

The Evidence-based summary generation stage is, therefore, logically, constrained by evidence. Descriptive statements report directly computed quantities, such as oxygen trends, hypoxia burden, heatwave days, stratification strength, or chlorophyll bloom frequency. Mechanism statements connect diagnostics only when the relevant diagnostics have been executed. Policy and management statements are framed as scientific decision support rather than legal, economic, or operational prescriptions. This design allows OceanMind to support scientific discovery and policy decision-making while preserving a transparent link between each conclusion and the computed evidence chain.

\subsection*{Reflection and replanning}

OceanMind uses execution feedback to revise workflows when the planned evidence chain cannot be completed. Failures fall into three main cases. In the first case, a registered tool fails at runtime. In the second case, a step generated by the CodingAgent returns an invalid artifact. Finally, in the third case, an intermediate result leaves a downstream dependency unresolved. For each case, the execution layer packages the failed step, error message, previous plan, and completed artifacts as runtime feedback. The PlannerAgent generates a repaired workflow only if the replanning budget remains and the original scientific intent can still be preserved.

Plan revision is bounded to avoid unlimited retry loops. A repaired workflow must pass the same validation used for the initial plan, covering tool names, parameter contracts, reference links, and data scope. Valid intermediate artifacts from earlier steps remain available to downstream steps, so successful computation is not repeated unnecessarily. If no valid repair is available, OceanMind will stop with an explicit failure message rather than produce an unsupported interpretation.

\subsection*{Graphical user interface (GUI) and streamed interaction}

OceanMind exposes the workflow through a web application. The backend is a FastAPI service for dataset metadata, manual visualization, query execution, streaming, and report export. The frontend is a Next.js application. It displays the route, planned workflow, step progress, result cards, visual evidence, source cards, and final synthesis. Manual visualization requests directly load a selected variable, region, time range, and vertical mode. Agent queries use the same data and rendering infrastructure after the workflow has been executed.

The streamed interaction also incorporates the MemoryAgent for use over multiple turns. Before planning, the MemoryAgent selects relevant memory from conversation, project, and reference stores. The selected packet is passed to the PlannerAgent. Before synthesis, the MemoryAgent selects memory that helps relate the current evidence to earlier context. For general answers, the MemoryAgent selects memory relevant to the user request without invoking dataset computation. After each completed turn, the MemoryAgent records reusable definitions, preferences, artifacts, and workflow patterns when appropriate. Explicit constraints in the current query always takes precedence over memory.

The streaming endpoint returns JSON events separated by newlines. During a dataset analysis turn, the interface first receives routing events, then memory events. Planning produces analysis proposals plus plan readiness events. Execution streams step starts, progress updates, step completions, warnings, and replanning events. Once results are available, the interface receives result events, synthesis readiness events, and the final response. Long tool calls emit heartbeat messages so that the interface remains responsive. Results are displayed as linked cards with map or chart panels before final synthesis is complete. This allows the user to inspect intermediate evidence while the interpretation is still being generated. Each completed conversation can also be exported as a report.

\subsection*{Benchmark and evaluation protocol}

We evaluated OceanMind with a dedicated benchmarking package. The benchmark contained 240 supported natural language queries derived from 59 target ocean skills (Supplementary Section~2.3). The benchmark combined 200 base queries with 40 interpretation queries. The suite spanned four complexity levels: L1 tested direct extraction, visualization and simple statistics (24 queries); L2 tested preprocessing, event detection, resampling and derived diagnostics (100 queries); L3 tested multi-step diagnostic composition, including sections, Hovm\"oller diagrams, EOF analysis, correlations and event comparison (84 queries); and L4 tested mechanism interpretation, budget reasoning, environmental assessment and evidence synthesis for policy questions (32 queries). Each query was paired with a canonical gold tool chain constructed from the current skill workflows and registered tool contracts.

We compared OceanMind with ReAct baselines implementing the original reason-act-observe paradigm~\cite{yao2023react}. In the main comparison shown in Fig.~\ref{fig:benchmarking}, OceanMind and ReAct(Claude) used Claude Sonnet 4.5 (\path{claude-sonnet-4-5-20250929}), whereas ReAct(Qwen) used Qwen3 Max (\path{qwen3-max}). In the backend-sensitivity evaluation shown in Supplementary Fig.~S2, OceanMind was separately evaluated with Claude Sonnet 4.5, Qwen3 Max, and DeepSeek V4 Pro (\path{deepseek-v4-pro}). All configurations received the same 240 queries, had access to the same 125 registered tools, and the same scoring protocol, and were allowed, at most, two execution-triggered repair or retry attempts.

Scoring applied the same rule-review gates to all benchmark traces. A run passed only when four gates were satisfied. The final deliverable had to match the requested result type. Execution had to reach a successful answer-bearing tool path; for OceanMind, the trace also had to include the required selector and planner stages. The executed tool calls had to preserve the variables, region, time window, depth selection and thresholds specified by the query. Finally, the completed diagnostics had to be sufficient for the expected skill. First-pass rate counted a passing run with no retry. Effective step efficiency was defined as $\min(N_\mathrm{ref}/N_\mathrm{run}, N_\mathrm{run}/N_\mathrm{ref})$, where \(N_{run}\) was the number of executed core tool calls and \(N_{ref}\) was the number of reference core tool calls in the gold chain. This metric penalized both missing and unnecessary core steps (Supplementary Section~2.2).

For the qualitative case comparisons in Fig.~\ref{fig:hypothesis-validation} and Supplementary Figs.~S3 and S4, identical natural language query text was supplied to the systems shown in each comparison, and we reported one recorded run per system. We collected the ChatGPT and OceanAI web responses on 3 August 2026, and we reproduced these responses here without manual rewriting. Sampling controls were not exposed by either web interface so they remained as the provider defaults.

\subsection*{Data availability}

The ocean state estimates we used in this study include CMOMS and CMEMS. The CMOMS data are available at \url{https://odmp.ust.hk/cmoms/}. CMOMS is a three-dimensional coupled physical-biogeochemical ocean state numerical model for the China Seas and adjacent western Pacific developed by the Ocean Dynamics and Modeling Program at the Hong Kong University of Science and Technology and has been validated and used in previous China Seas studies~\cite{gan2016numerical,gan2016three,liu2017pathways,gan2022hotspots}.

The CMEMS data we used are publicly available from the Copernicus Marine Data Store (\url{https://data.marine.copernicus.eu/}) and were extracted using the Copernicus Marine Toolbox. For the Gulf of Mexico and northern Caribbean application, we obtained the physical fields from the Global Ocean Physics Analysis and Forecast product  (product ID \path{GLOBAL_ANALYSISFORECAST_PHY_001_024}; DOI: \url{https://doi.org/10.48670/moi-00016}) using dataset IDs \path{cmems_mod_glo_phy-thetao_anfc_0.083deg_P1D-m} for temperature (\path{thetao}), \path{cmems_mod_glo_phy-so_anfc_0.083deg_P1D-m} for salinity (\path{so}), and \path{cmems_mod_glo_phy-cur_anfc_0.083deg_P1D-m} for zonal and meridional currents (\path{uo}, \path{vo})~\cite{copernicusGlobalPhy2026}. We obtained the biogeochemical fields were obtained from the Global Ocean Biogeochemistry Analysis and Forecast product (product ID \path{GLOBAL_ANALYSISFORECAST_BGC_001_028}; DOI: \url{https://doi.org/10.48670/moi-00015}) using \path{cmems_mod_glo_bgc-pft_anfc_0.25deg_P1D-m} for chlorophyll-a (\path{chl}) and \path{cmems_mod_glo_bgc-bio_anfc_0.25deg_P1D-m} for dissolved oxygen (\path{o2})~\cite{copernicusGlobalBGC2026}. These subsets covered 98.68\(^{\circ}\)W-76.98\(^{\circ}\)W, 16.35\(^{\circ}\)N-32.02\(^{\circ}\)N, 1 April 2025 to 31 March 2026 and at depths from 0.494 to 5727.917~m. For the tropical Pacific SST application, we obtained \path{thetao} from the Global Ocean Physics Reanalysis (product ID \path{GLOBAL_MULTIYEAR_PHY_001_030}; dataset ID \path{cmems_mod_glo_phy_my_0.083deg_P1D-m}; DOI: \url{https://doi.org/10.48670/moi-00021})~\cite{copernicusGlobalPhyReanalysis2026}. This subset covered 120\(^{\circ}\)E-80\(^{\circ}\)W, 10\(^{\circ}\)S-10\(^{\circ}\)N from 1 January 2005 to 31 December 2025. We accessed the  CMEMS product catalogue records and identifiers were accessed on 4 August 2026.

The SST reference fields we used for the independent surface comparison were derived from the GHRSST Level 4 MUR Global Foundation Sea Surface Temperature Analysis, version 4.1 (\texttt{MUR-JPL-L4-GLOB-v4.1}), produced by the JPL MUR MEaSUREs Project and distributed by NASA PO.DAAC. The source dataset is available from PO.DAAC (\url{https://podaac.jpl.nasa.gov/dataset/MUR-JPL-L4-GLOB-v4.1}; DOI: \url{https://doi.org/10.5067/GHGMR-4FJ04}), and the Zarr access route used by the processing script is available through the AWS Open Data Registry (\url{https://registry.opendata.aws/mur/}; \url{https://mur-sst.s3.amazonaws.com/zarr-v1/analysed_sst/}). The field observations we used for in situ verification are the \texttt{201801\_XMU\_CHOICE-II} CTD downcast profiles. These data are available from the corresponding author (J.G.; \href{mailto:magan@ust.hk}{\texttt{magan@ust.hk}}) upon reasonable request.

\section*{Code availability}
The OceanMind source code is available at \url{https://github.com/YangLabHKUST/OceanMind}. And our website is \url{http://oceanmind.wavyocean.hkust.edu.hk/}. If your browser has trouble accessing the site over HTTP, try opening this address directly: \url{http://143.89.31.83}. The record of the benchmark and evaluation protocol is available at \url{https://doi.org/10.5281/zenodo.21189249}.

\section*{Acknowledgements}

The work described in this paper was fully supported by a grant from the Research Grants Council of the Hong Kong Special Administrative Region, China (Project Reference Number: AoE/P-601/23-N); we acknowledge the support of the Center for Ocean Research in Hong Kong and Macau (CORE). CORE is a joint research center for ocean research between Laoshan Laboratory and HKUST.

\section*{Author contributions}

F.Z., C.Y. and J.G. conceived the study and designed the OceanMind framework. Y.C, H.K, Y.Z., A.H. and Q.Z. provided valuable insights for framework design. F.Z. implemented OceanMind and evaluated the baseline methods. F.Z. and W.C. processed the ocean data and interpreted the results. F.Z., C.Y. and J.G. wrote the manuscript. All authors reviewed and approved the manuscript.

\section*{Competing interests}

The authors declare no competing interests.

\bibliographystyle{unsrt}
\bibliography{refs}

\

\end{document}